\documentclass[useAMS,usenatbib]{mn2e}
\usepackage[cp1251]{inputenc} 
\usepackage{graphicx}
\usepackage{amssymb}
\usepackage[english]{babel}
\usepackage{captcont}
\usepackage{morefloats}
\usepackage{datetime}
\usepackage{afterpage}
\usepackage[pdftex]{hyperref}
\usepackage{ulem} 

\usepackage{longtable}
\usepackage{lscape}
\usepackage{afterpage}
\usepackage{booktabs,fixltx2e}
\usepackage[flushleft]{threeparttable}

\usepackage[dvipsnames]{color}

\def\ltsima{$\; \buildrel < \over \sim \;$}
\def\simlt{\lower.5ex\hbox{\ltsima}}
\def\gtsima{$\; \buildrel > \over \sim \;$}
\def\simgt{\lower.5ex\hbox{\gtsima}}

\newcommand{\kmskpc}{km~s\ensuremath{^{-1} }~kpc\ensuremath{^{-1} }}

\newcommand{\Msun}{M\ensuremath{_\odot}}
\newcommand{\Oo}{\displaystyle}

\title[Dyncamical friction on GMCs in M~33]{Giant molecular clouds in M~33: are they susceptible to dynamical friction?}

\author[Zasov A.V. and Khoperskov S.A.]{A.V. Zasov$^{1,2}$\thanks{zasov@sai.msu.ru} and S.A. 
Khoperskov$^{3,4,2}$ \\ $^1$Physical department of the Moscow M.V. Lomonosov State University, Leninskie gory 1, Moscow 119991, Russia  \\
$^2$Sternberg Astronomical Institute, 
Moscow M.V. Lomonosov State University, 
Universitetskij pr., 13,  Moscow, 119992, Russia \\ $^3$Dipartimento di Fisica, Universit\`{a} degli Studi di Milano, via Celoria 16, I-20133 Milano, Italy \\ $^4$Institute of Astronomy, Russian Academy of 
Sciences, Pyatnitskaya st., 48, 119017 Moscow, Russia}

\date{\today}

\pagerange{\pageref{firstpage}--\pageref{lastpage}} \pubyear{2015}

\def\LaTeX{L\kern-.36em\raise.3ex\hbox{a}\kern-.15em
    T\kern-.1667em\lower.7ex\hbox{E}\kern-.125emX}

\begin{document}

\maketitle

\begin{abstract}
Most of giant molecular clouds~(GMCs) in M~33 are connected  with spiral-like gaseous arms (filaments) with the exception  of the inner 2~kpc region where the link between the arms and GMCs disappears~\citep[see][]{2011PASJ...63.1171T}. We check whether it may be caused by the dynamic friction retarding the clouds. Using  semi-analytical model for this galaxy we calculate the dynamics of GMCs of different masses situated at different initial galactocentric distances in the disk plane. We demonstrate that the dynamical friction may really change the orbits of GMCs in the central 2 kpc-size region. However in this case the typical lifetimes of GMCs should be close to or greater than $10^8$~yr, which is larger than the usually accepted values.	
\end{abstract}


\section{Introduction}
Giant molecular clouds  (GMCs) are the most massive bodies in galactic discs. The physical process of their  formation from the more rarefied gas is the matter of debate, although  it seems evident that there  is no universal  mechanism of transition of HI to H$_2$. It may happen either through gas flow 
collisions, or by the process of slow contraction of a cloud from diffuse atomic or 
molecular medium~\citep[see, e.g.][]{2013MNRAS.432..653D}. 
Observations clearly show that in the case of contrast spiral arms such as observed in  M~51, GMCs 
are concentrated in the arms, although less massive clouds are dispersed both along 
the arms and in between, giving evidence that molecular clouds may survive when passing through 
spiral arms~\citep[see review by][]{2013ASPC..476...49K}. In the low luminous galaxies such  as LMC, SMC or M~33 the situation is different:  GMCs are usually connected with HI filaments which means that most of their lifetime they move parallel with HI. How long do they live is another question.  It is usually accepted that GMCs have a relatively short lifetime $~10^6-3\cdot 10^7$ yr (several free-fall times) and  are disrupted by stellar feedback soon after the beginning of star formation~\citep{2013MNRAS.432..653D, 2014MNRAS.437L..31D,2011ApJ...729..133M}. 
However there are arguments that at least a fraction of GMCs in galaxies may live as long as about  $~10^8$ yr or more~\citep[see discussion in][]{2013ASPC..476...49K,2013seg..book..491S,2014Ap&SS.353..595Z}. It is worth noting that the feedback effectiveness of disrupting of a cloud is badly known. Numerical  simulations  of evolution of GMCs demonstrate that the gas outflows from the feedback may reduce the mass of the cloud but do not destroy it in case of the weak thermal feedback \citep{2015ApJ...801...33T}. However another numerical simulations (e.g. by \citet{2014MNRAS.442.3674W} show that GMCs could be destroyed by the stellar feedback.

 It is evident that the dynamical evolution of GMCs depends on their existence as 
single entities. In the case of their long lifetime one can expect that the dynamical friction may be unavoidable in the inner part of a disc leading to the radial migration of gas toward the centre of a galaxy. This scenario was first proposed by ~\citet{1987Afz....26..443S}. Later this idea was applied to the Milky Way in the frame of 2D analytical  model of MW by \citet{1990PASJ...42..517Y}. In turn \citet{1999ApJ...526..665J} proposed the dynamic friction of very massive gas clumps observed in the bar of starburst galaxy NGC~2782 as the mechanism forcing the gas to inflow to the centre of this galaxy. In principle, dynamical friction may be responsible for central gas concentration and stimulation of starforming or AGN activities in some gas-rich galaxies.

Slowly  rotating galaxy with  low contrast spiral structure such as M~33 may be a good laboratory to test the dynamical evolution of GMCs. \citet{2012ApJ...761...37M} estimated a lifetime of GMCs in M~33 by comparing CO(J=3-2) data and young star ages. They concluded that about 2/3 of all GMCs have the age $\le10^7$~yr, although the age of the rest clouds is $10-30$ Myr. However one should have in mind that these estimates are not too reliable being based on the visible manifestation of star formation inside of the clouds. In addition the starting point of the detachment of GMCs from the surrounding gas remains rather indefinite \citep{2013MNRAS.432..653D} especially  if taken into account that a significant part of relatively low-dense translucent gas in the outer regions of slowly forming cloud may be non-detected in CO-lines.

Most of GMCs in M~33 are linked with the gaseous spiral-like arms. However  in the inner 2 kpc region of this galaxy the tight connection between GMCs and HI filaments disappears ~\citep{2011PASJ...63.1171T}, and concentrations of molecular gas look quite detached from HI  there. As it was noted by~\citet{2013ASPC..476...49K}, discussing this result, ``GMCs are decoupled from the HI distribution as if they are entities that survive through almost a galactic rotation period". It is worth trying to check whether this decoupling may be explained by the dynamical friction which  brakes GMCs moving through the stellar disk and the bulge and forces them to depart from their birthplaces in the gaseous arms. Of course, the result depends on the ratio of the braking time to the typical  life time of GMCs.  We have developed a numerical model of motion of point-like GMCs  in the galactic disc of M~33 and calculated the drag forces which affect the clouds within the radial distance of several ~kpc. We argue that the spatial offset between GMCs and their  birth places (spiral-like filaments) may be accounted for dynamic friction   if the lifetime of GMCs is allowed to exceed the period of rotation in the inner disc~($\sim 10^8$ yr). 

The paper is organized as the following. The second Section contains a basic model description including both the equation of cloud motions and the initial parameters of simulations. In the third Section the main results are described. The last Section contains the discussion and summary.

\begin{figure}\label{fig::rotation}
\includegraphics[width=1\hsize]{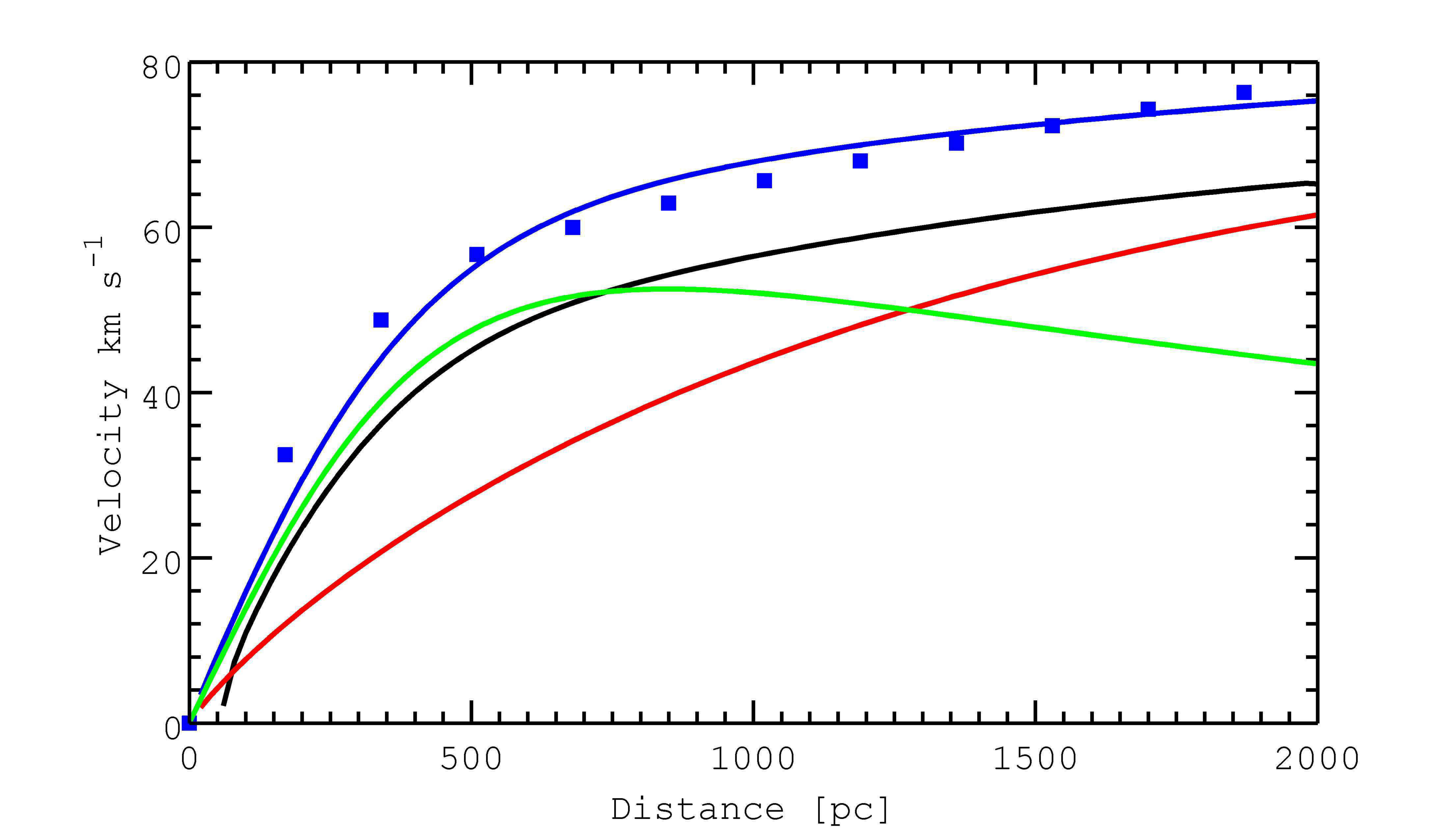}
\caption{Model fit of the rotation curve decomposition for M~33 for the inner r = 4 kpc region: bulge~(green) and stellar disc~(red) circular velocities. Gas circular motion is represented by blue line. Stellar disc rotation curve is shown by black line. Points represents the observational data following by~\citet{2003MNRAS.342..199C}. } 
\end{figure}

\section{The Model}
We designed a simple kinematics model of the GMCs motion through the stellar 
components of the galaxy. A cloud is considered as a particle $m_c$ moving through the sea of ''stellar'' particles of mass~$m* \ll m_c$ whose spatial distribution reproduces the spherical bulge component and the galactic disc.   As a basic model, we used the model presented by~\citet{2013AN....334..785S}, which is based on the rotation curve of M~33 and stellar velocity dispersion of disc stars, where the density of disc was calculated using the condition of its marginal stability. Similar density distribution of a  stellar disc was found in the more sophisticated mass model  developed by~\citet{2014A&A...572A..23C}.  Although  there is no strict evidences of the classical bulge  in M~31~\citep[see discussion in][]{2007ApJ...669..315C},   ~\citet{1994ApJ...434..536R} found the excess of  spherically distributed light fitted by $r^{1/4}$ law  with the effective radius  $\approx 8$~arcsec $\approx  2$~kpc, which may be attributed to the bulge or inner stellar halo. This excess fits the shape of rotation curve in the inner region of M~33~\citep{2012AstL...38..139S}. The only change we have done in our calculations  is the replacement of de Vaucouleurs bulge by Plummer sphere with mass $10^8$~$M_\odot$ and scale length 0.6~kpc. Note however that the resulting force of dynamical friction is determined mostly by the disc (see below), being not sensitive to the bulge parameters. The exponential stellar disc is characterized by the central surface density 800~$M_\odot$pc$^{-2}$, radial scale $1.9$~kpc and vertical scale length $200$~pc. Stellar velocity dispersion in the model corresponds to the disc scale height.

 As it is shown in  Fig.~\ref{fig::rotation}, our model satisfactory reproduces the velocity curve of the inner part of the galaxy. We do not consider the dark halo separately, but it is worth noting that its relative  mass is low in the considered region, and,  moreover, the role of dark matter is inseparable from the role of bulge-like spherical component.  We assume the rigid non-evolved model of the spherical component and galactic disc that allows to avoid the direct self-consistent N-body simulations. 
 
 To describe a dynamics of massive particles (clouds) we have used a several well-known relations relevant to the dynamical friction. The equation of motion has the form:
\begin{equation}
\Oo m_c \frac{\partial^2\bf r}{\partial t^2} + \beta \frac{\partial \bf r}{\partial t} + 
\frac{V^2_c}{r^2}{\bf r} = 0\,,\label{eq::dyn1} 
\end{equation}
where ${\bf r}$ is a cloud radius vector, $\beta = \beta_d + \beta_b$ is a drag coefficient by the disc and spherical components and $V_c$ is a circular velocity supported by the gravitational potential of these components.

\begin{figure*}
\includegraphics[width=0.195\hsize]{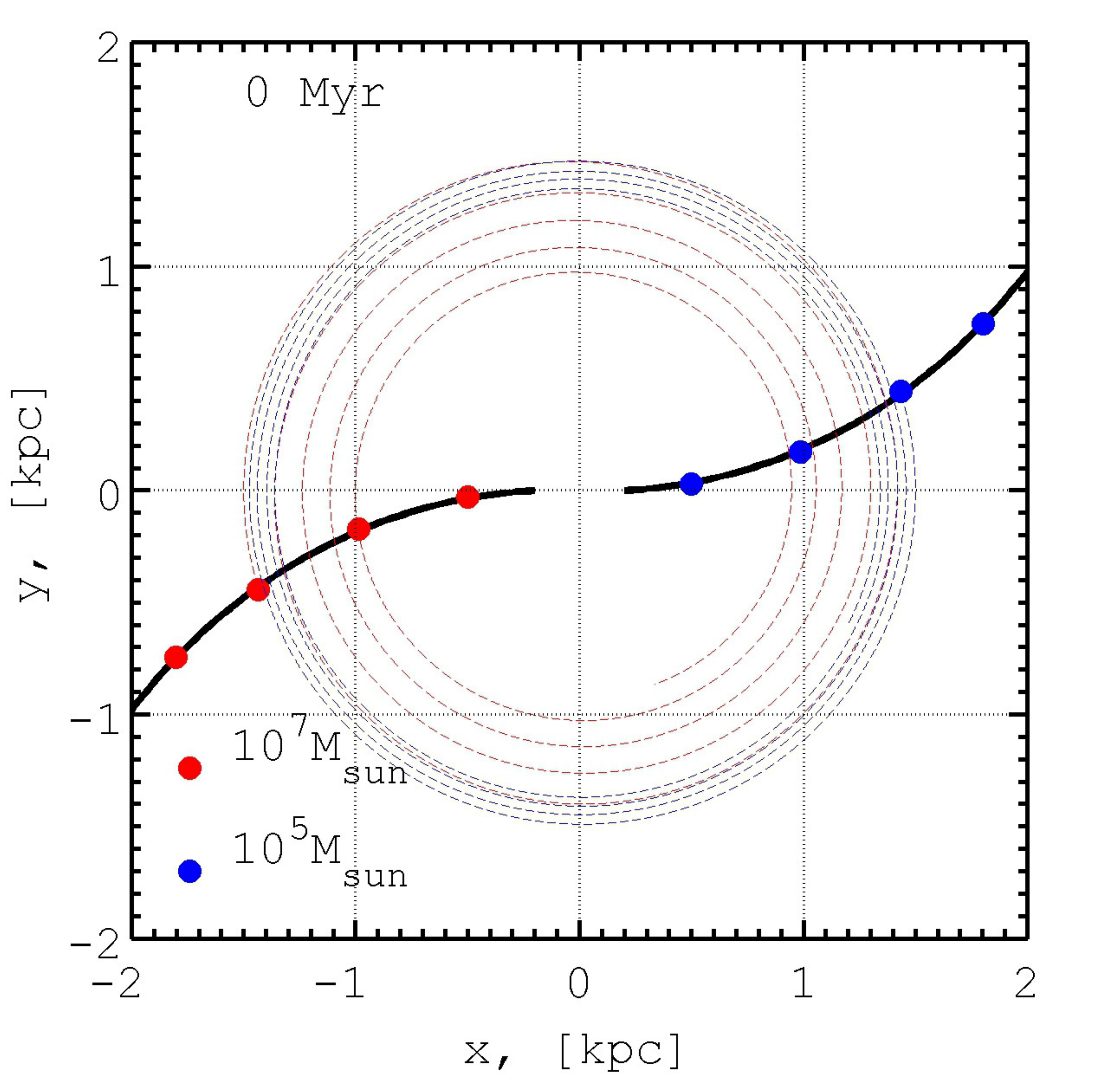}
\includegraphics[width=0.195\hsize]{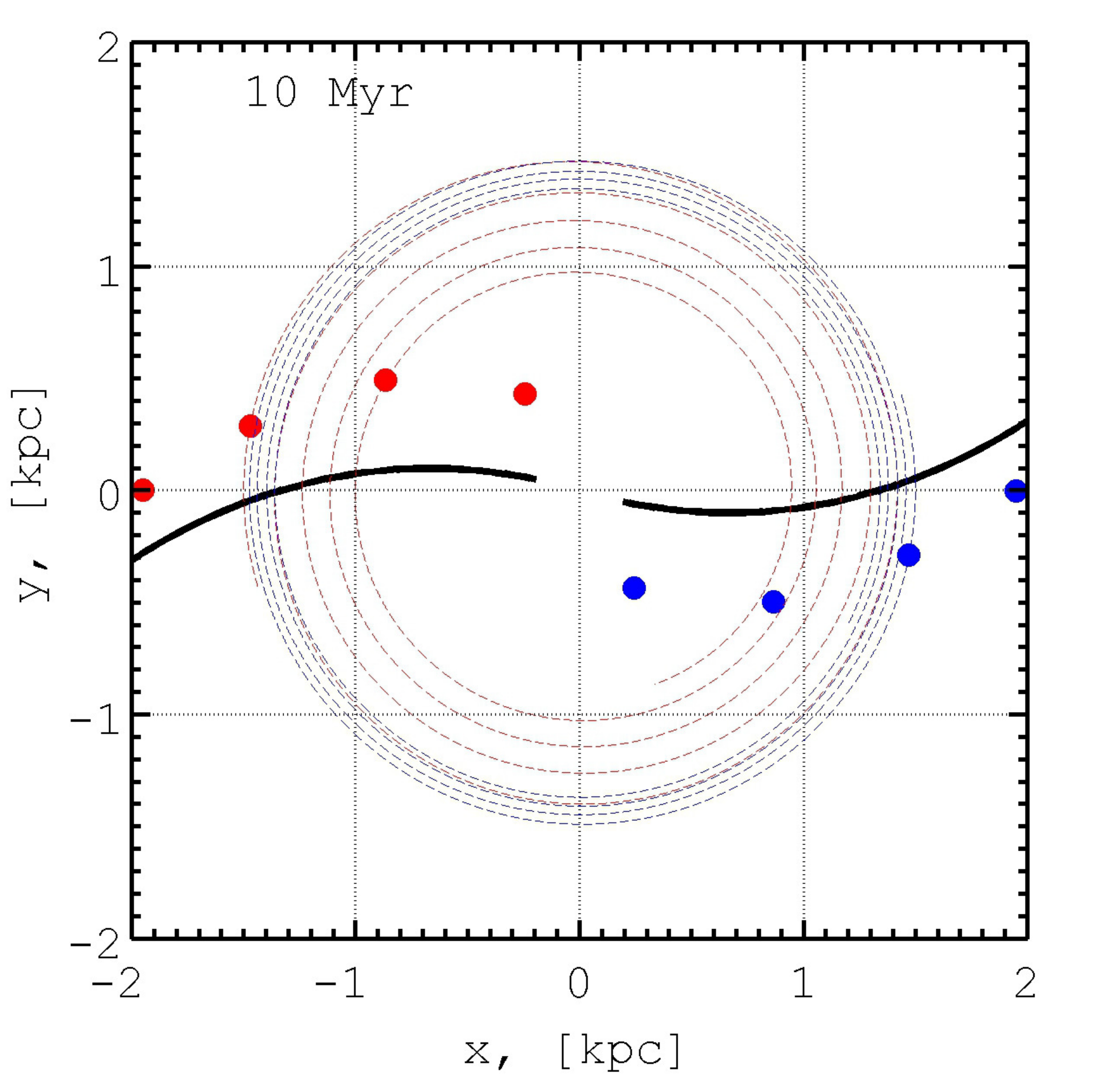}
\includegraphics[width=0.195\hsize]{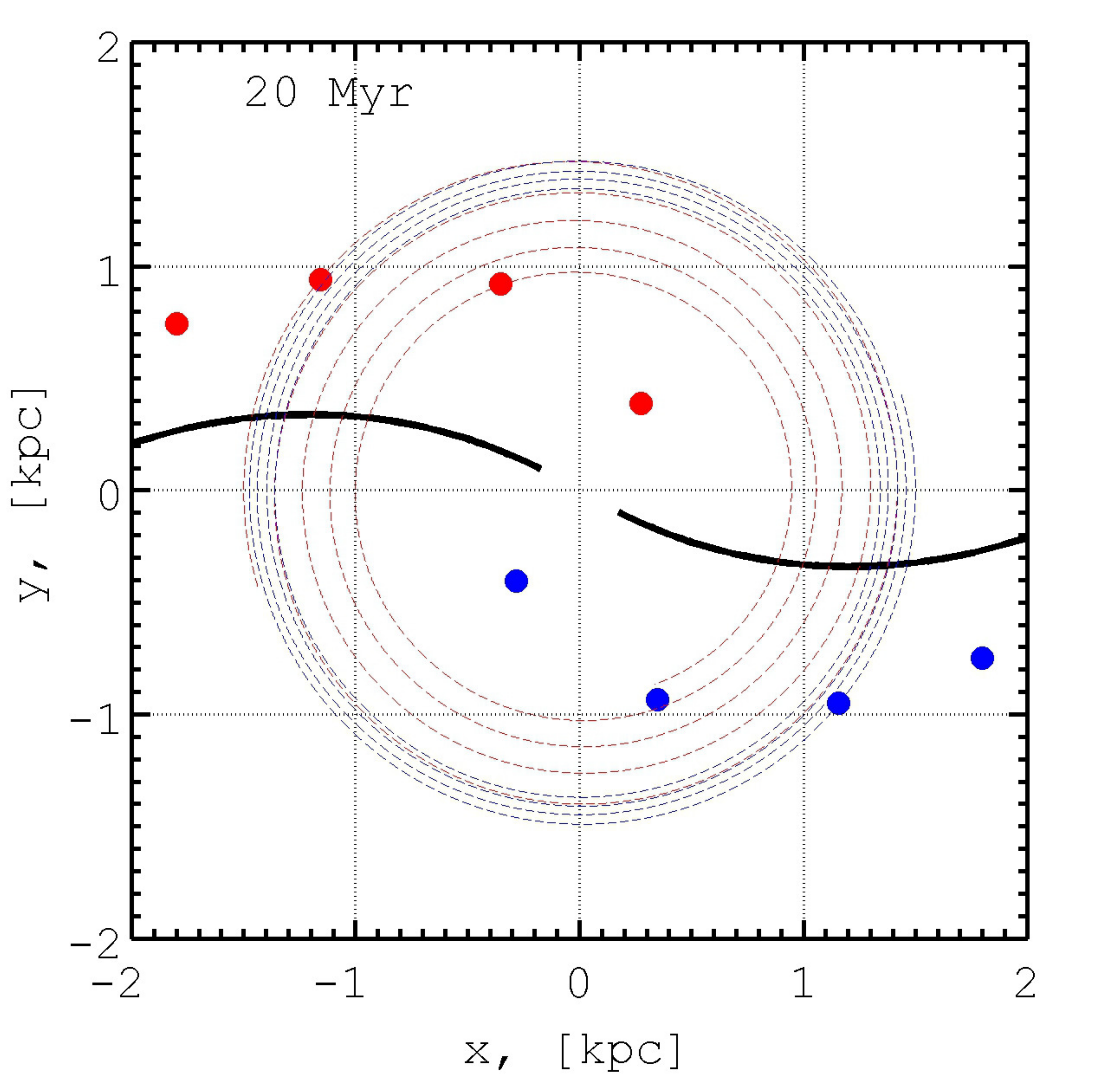}
\includegraphics[width=0.195\hsize]{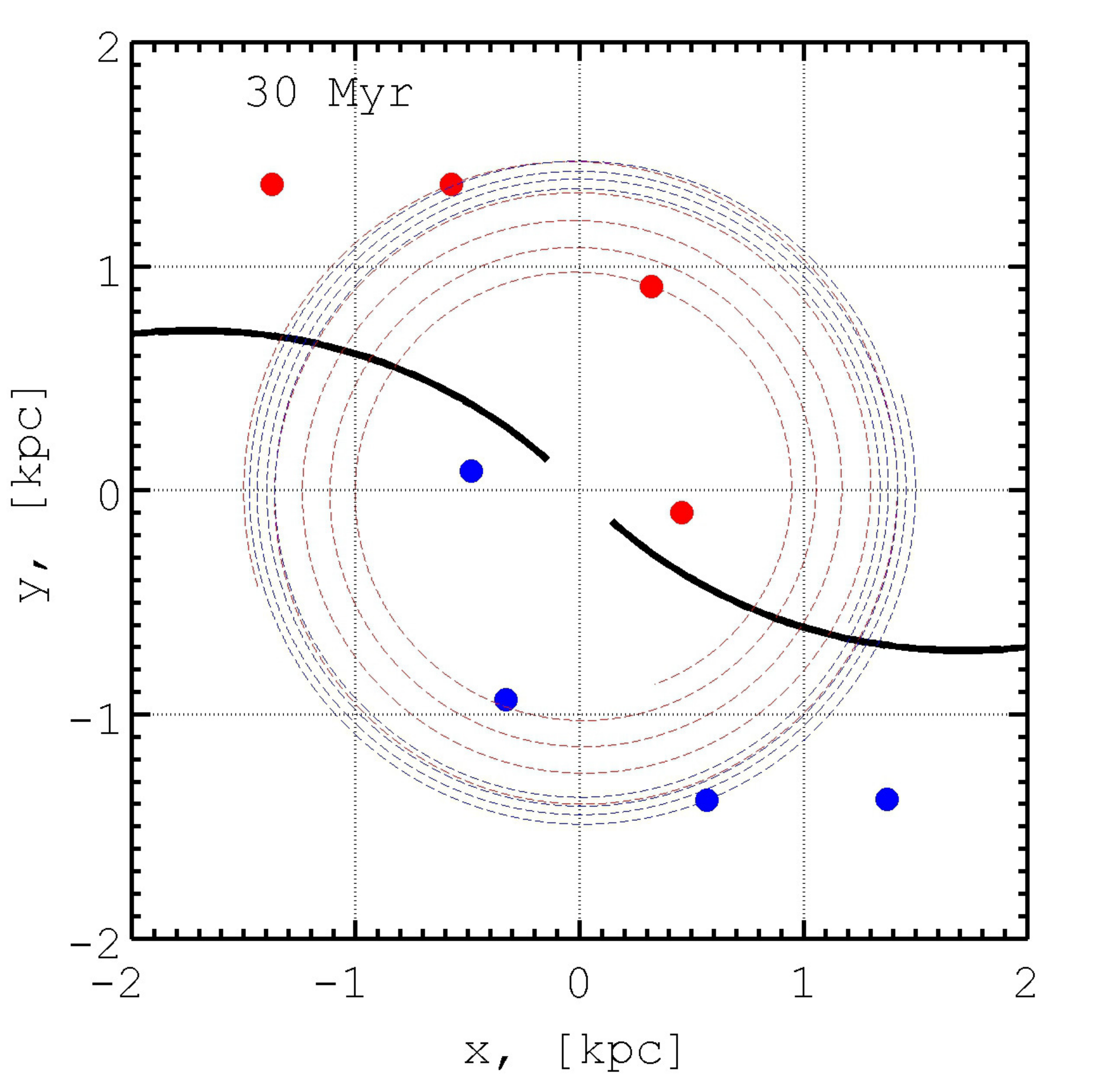}
\includegraphics[width=0.195\hsize]{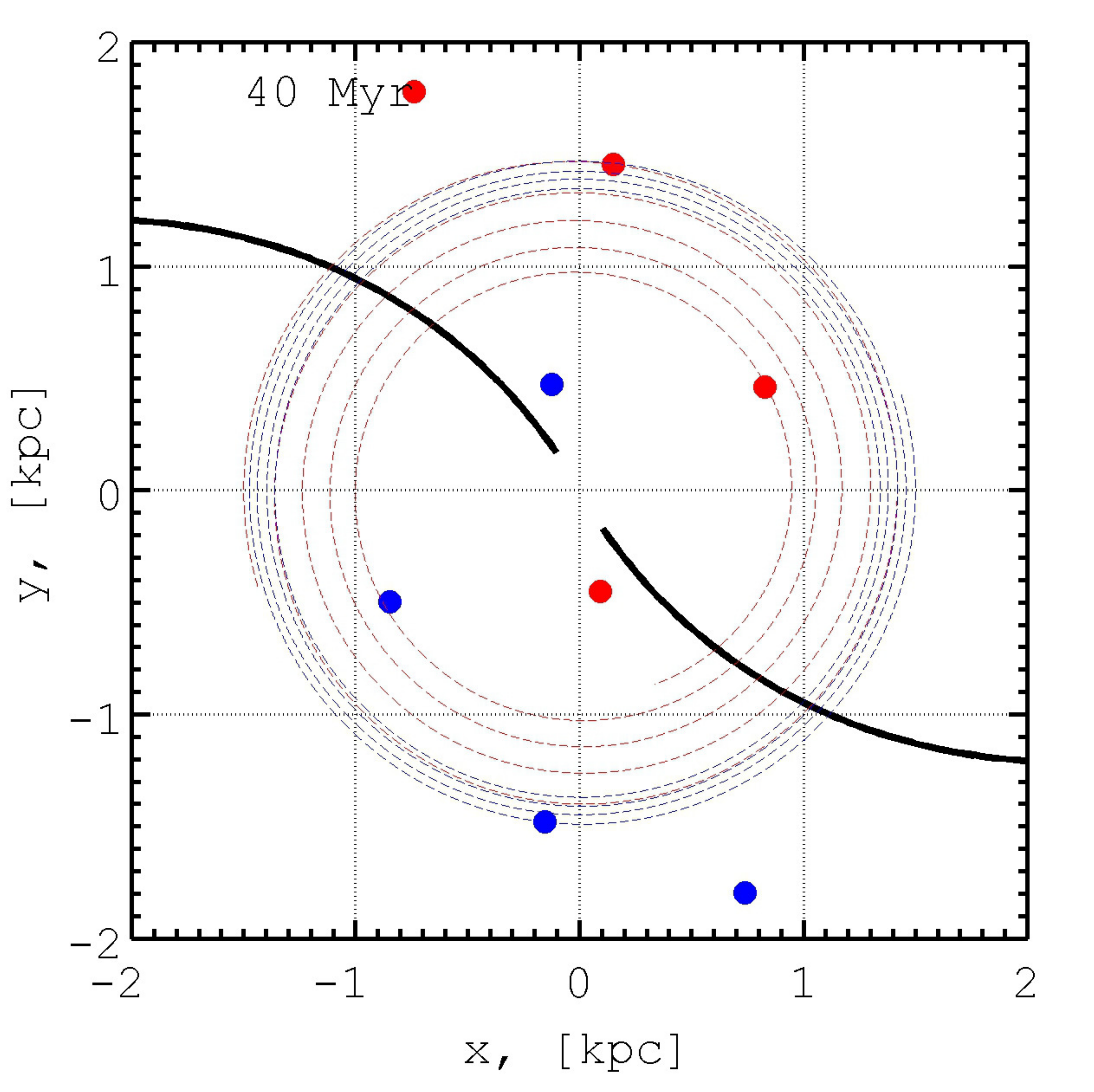}
\includegraphics[width=0.195\hsize]{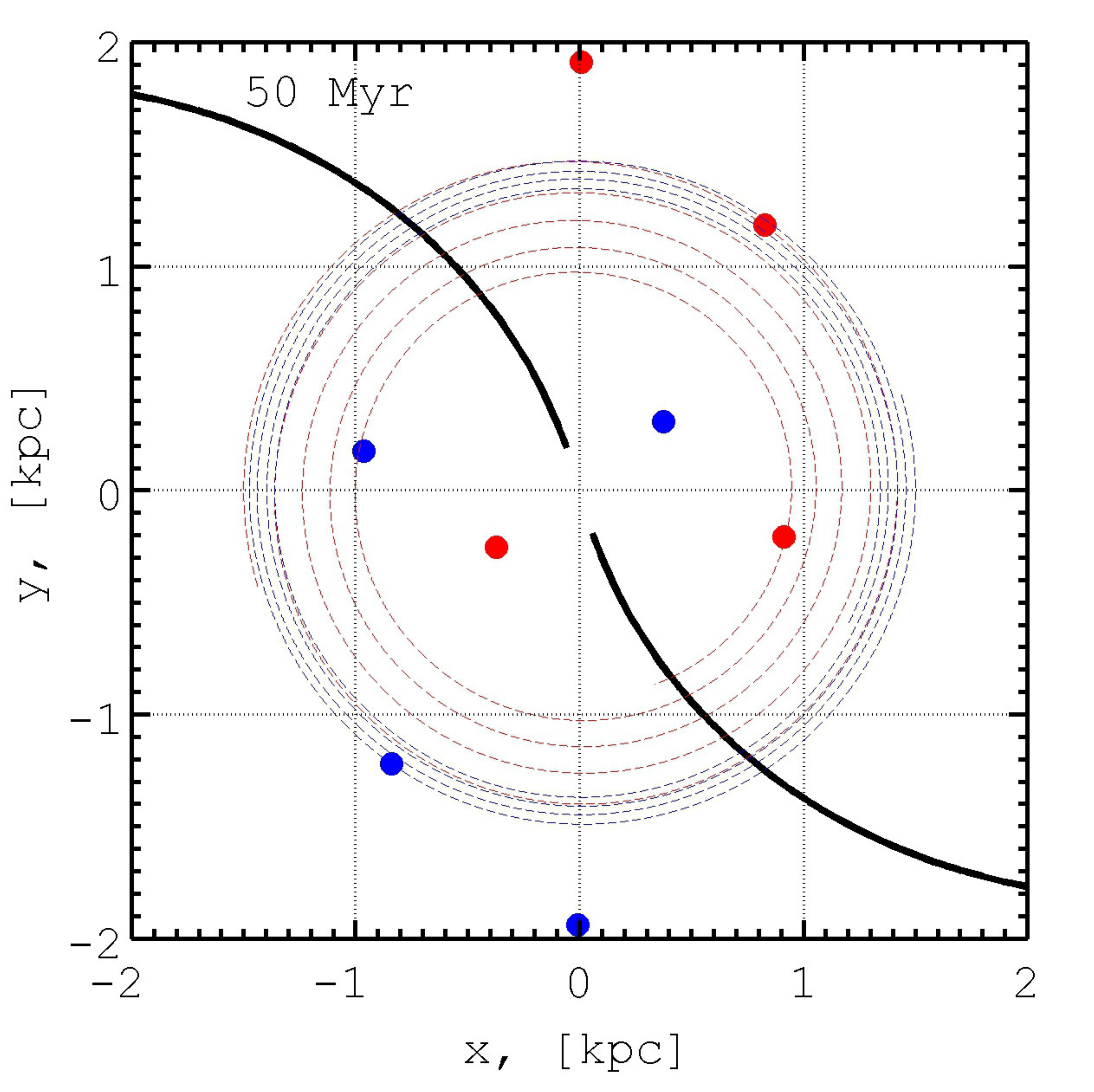}
\includegraphics[width=0.195\hsize]{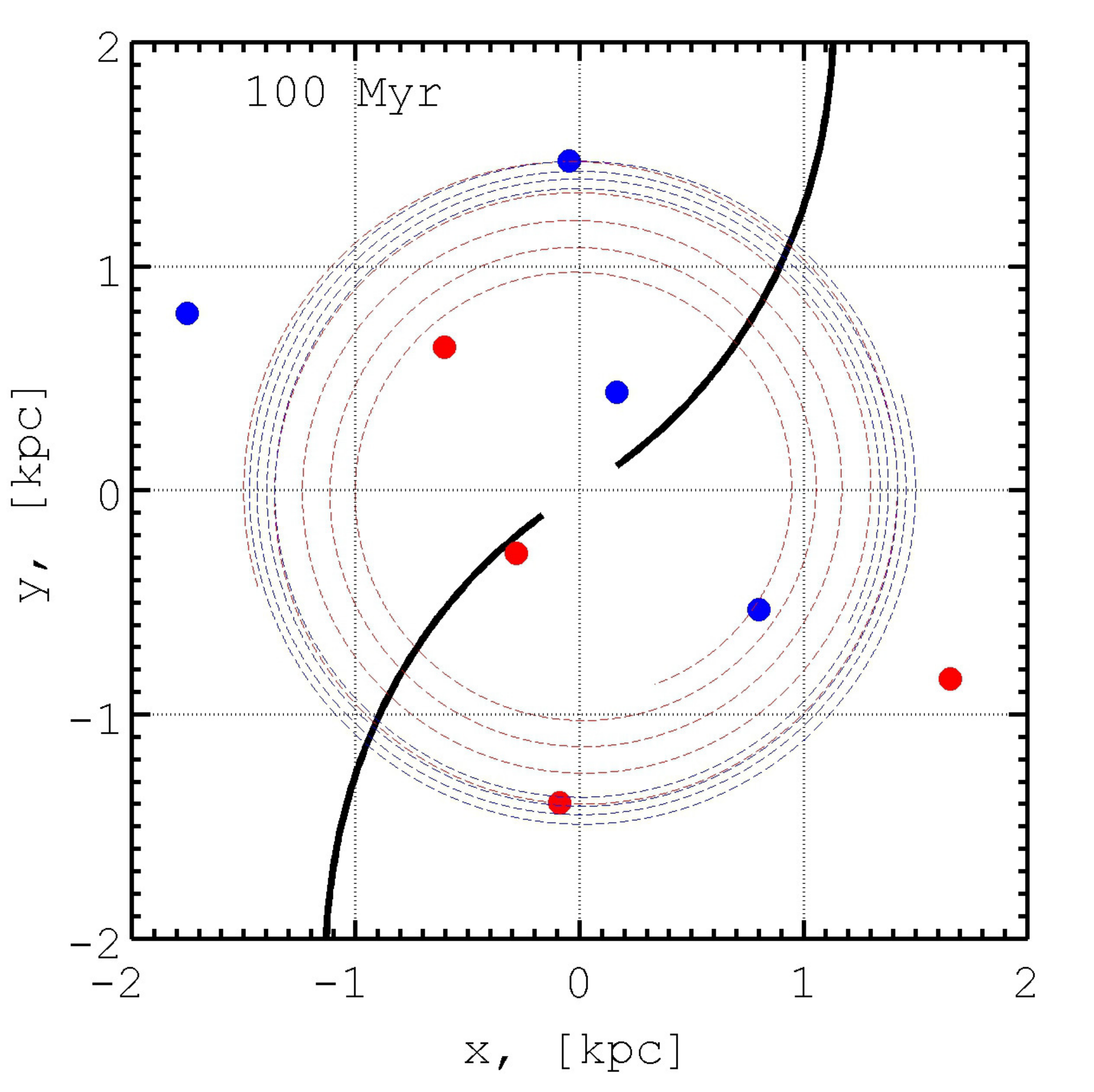}
\includegraphics[width=0.195\hsize]{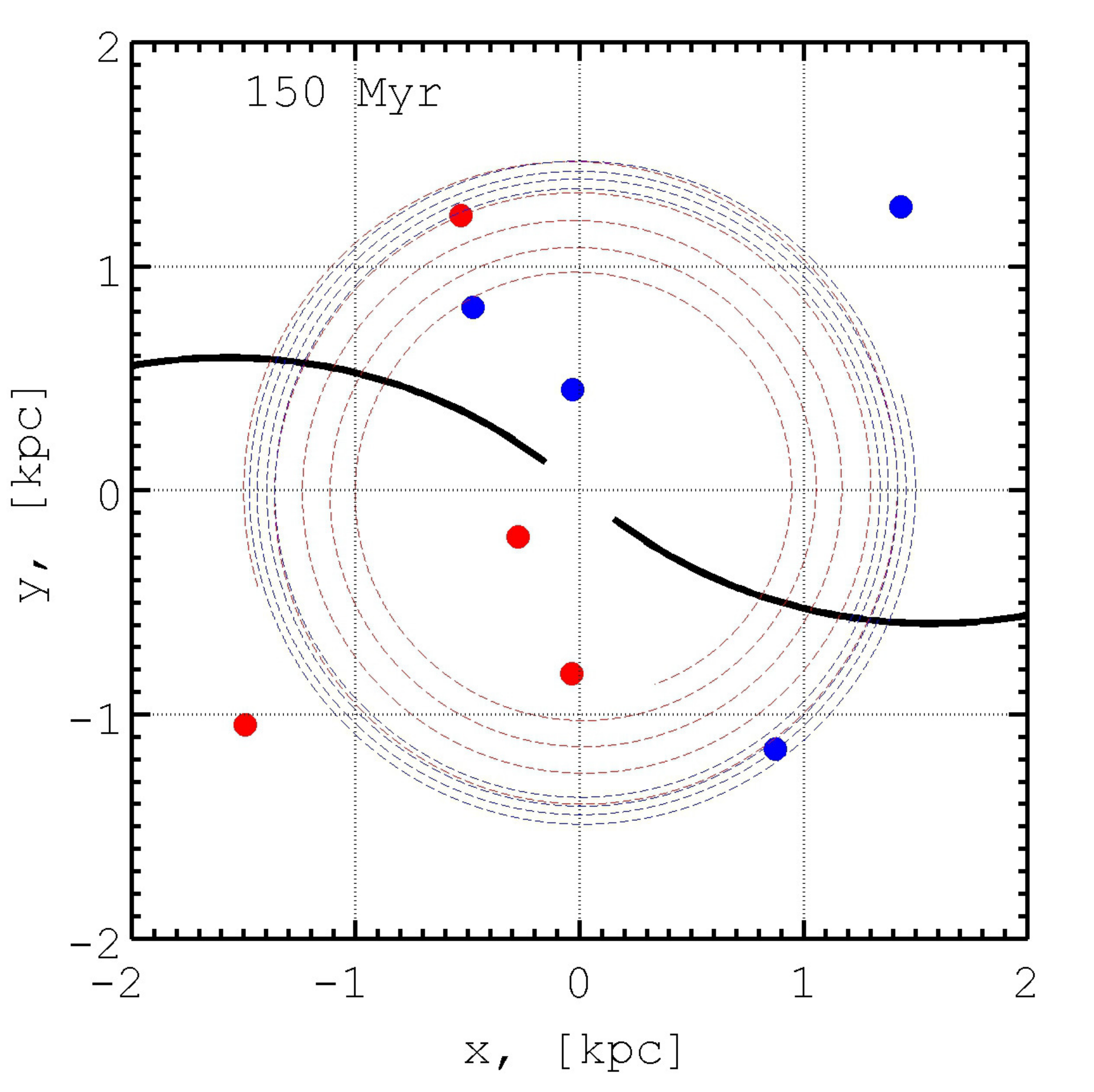}
\includegraphics[width=0.195\hsize]{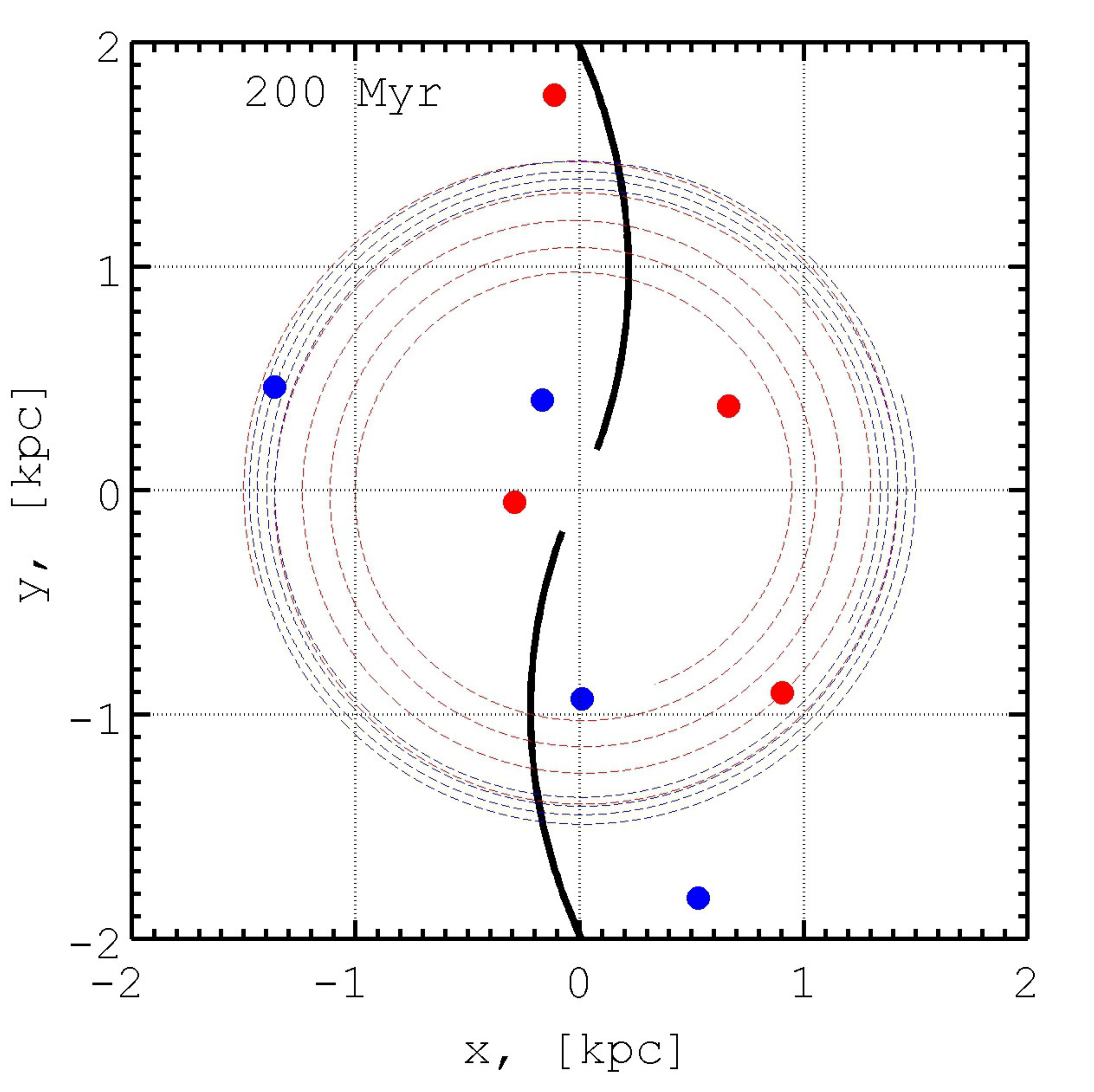}
\includegraphics[width=0.195\hsize]{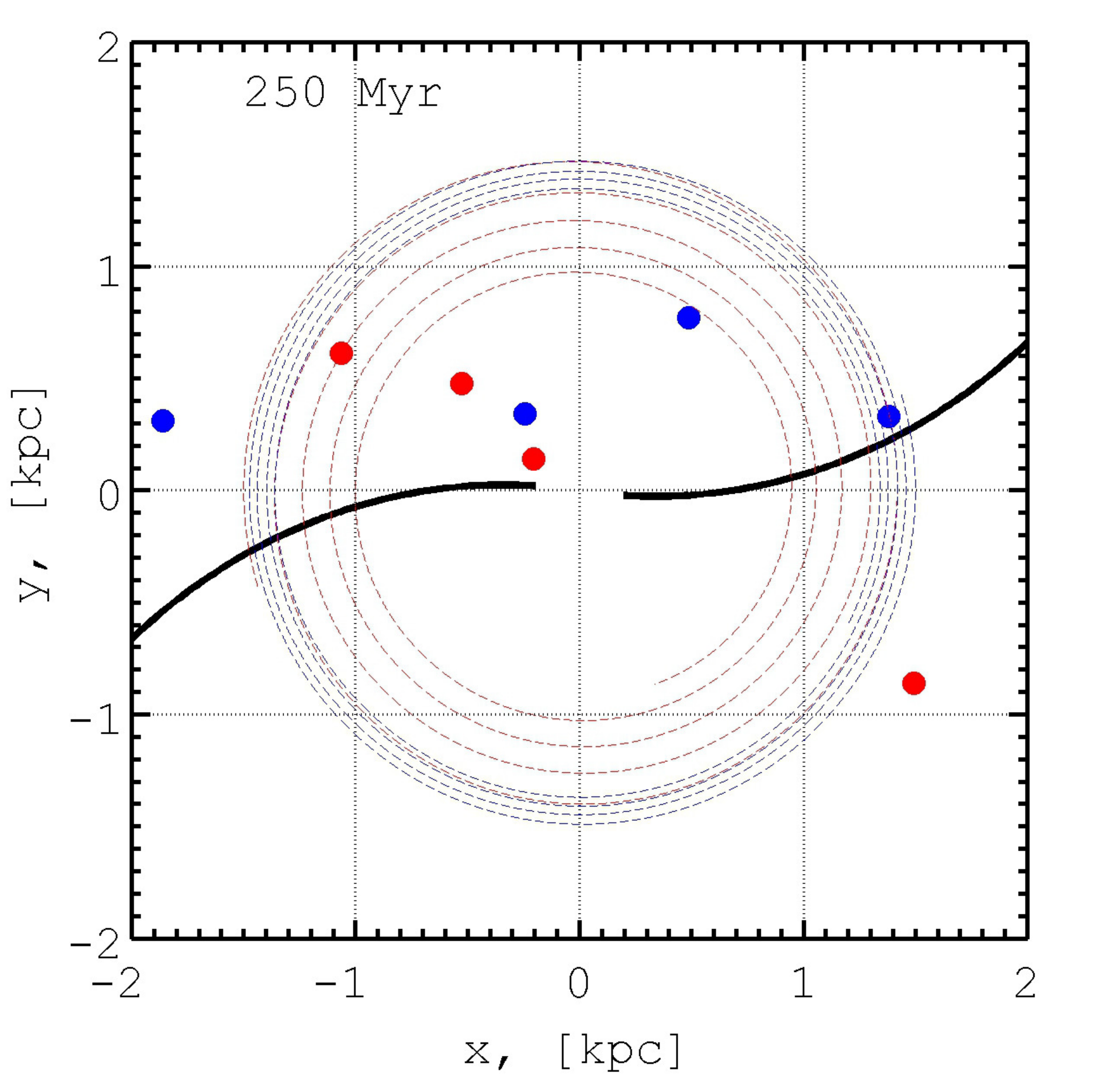}
\caption{ Time sequence of the cloud positions for the clouds of $10^5$\Msun~(blue) and $10^7$\Msun~(red) in the fixed reference system. Black lines mark a spiral structure rotating rigidly with an angular velocity $25$~km~s$^{-1}$~kpc$^{-1}$. Dashed lines are the cloud tracks beginning from the initial position $r_0 = 1500$~pc.}\label{fig::evol}
\end{figure*}

\begin{figure*}
\includegraphics[width=0.49\hsize]{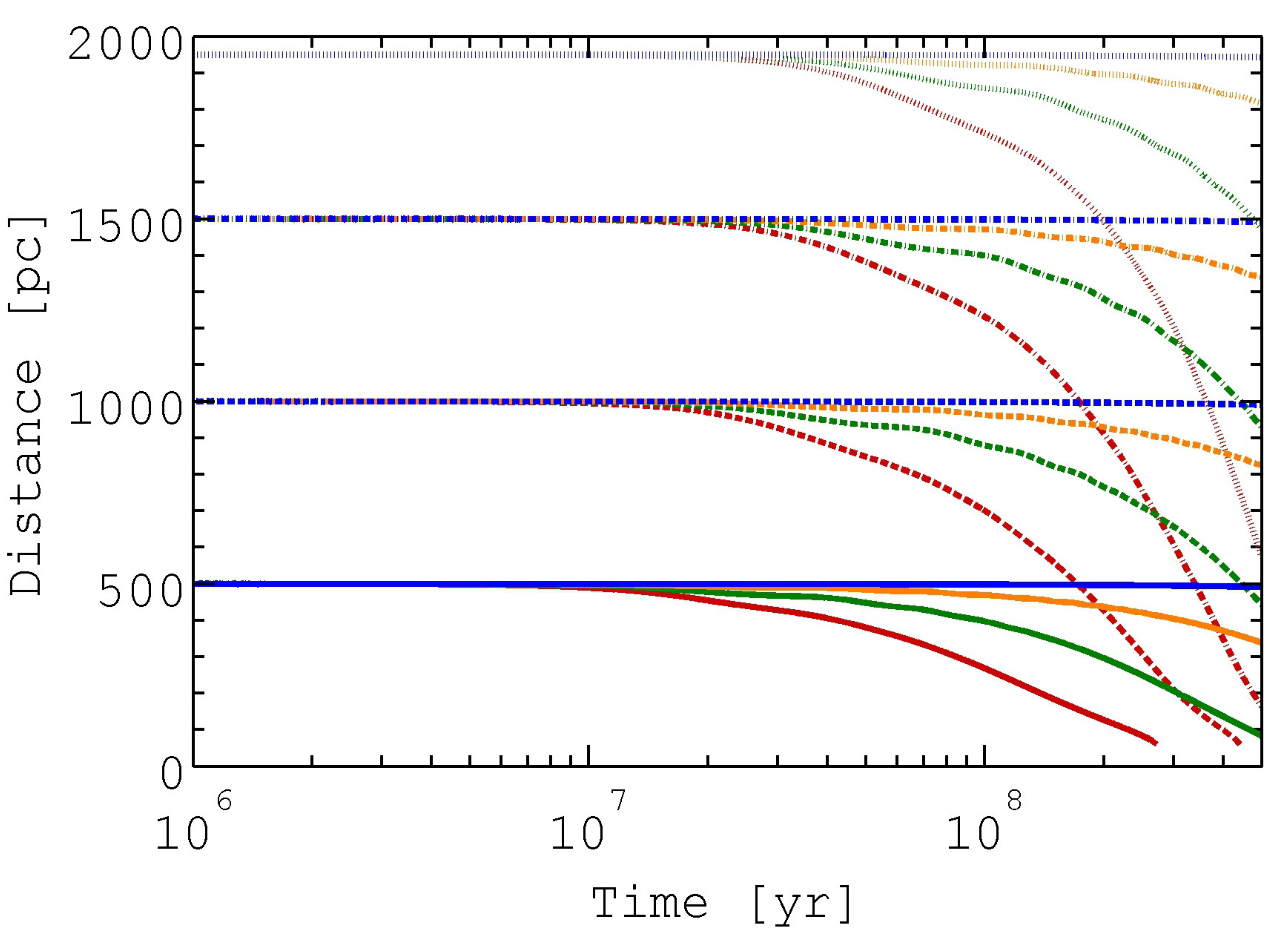}
\includegraphics[width=0.49\hsize]{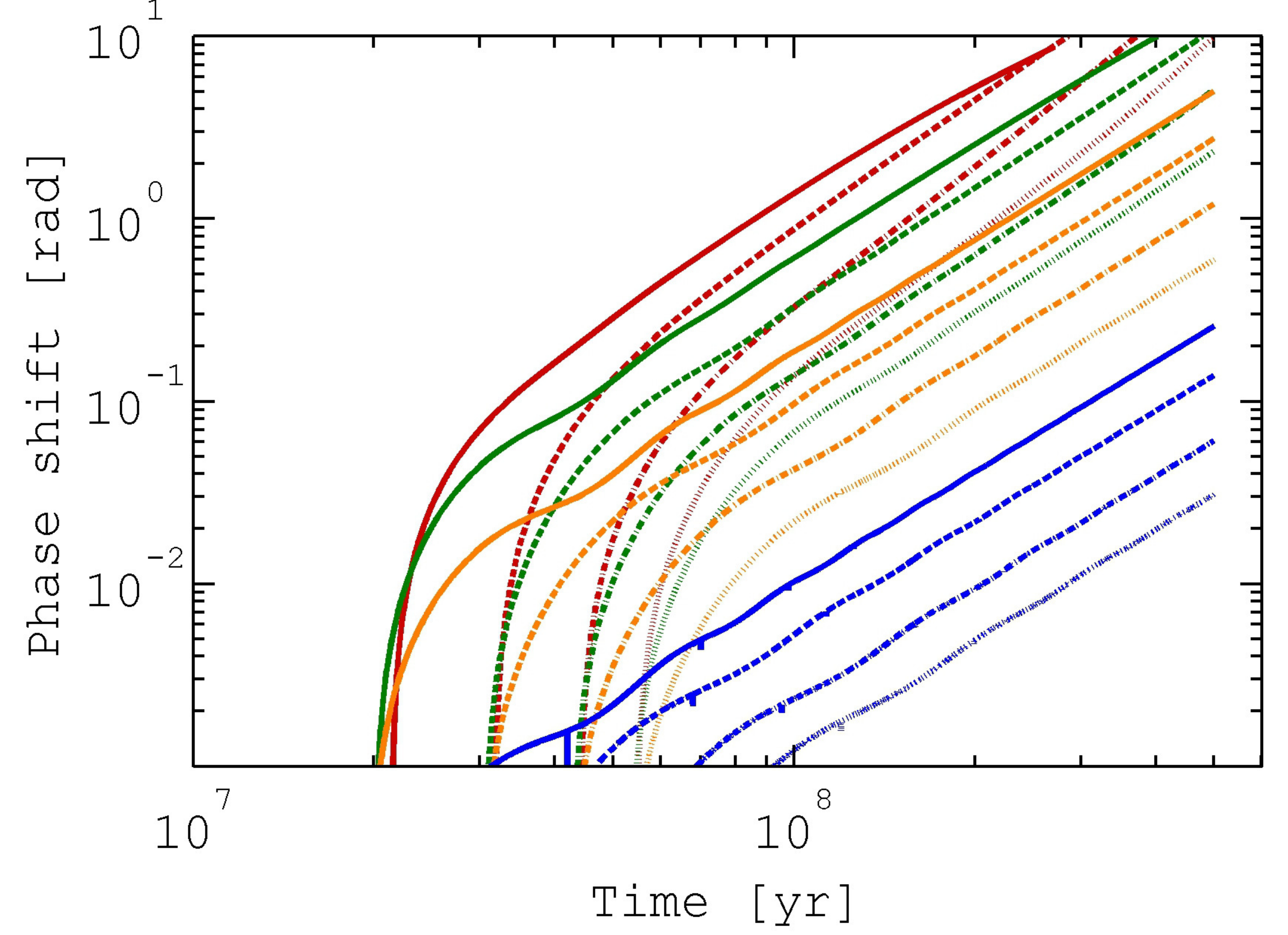}
\caption{Left:  evolution of galactocentric distance $r$ for clouds with masses 
$10^4$~(deep blue), $10^5$~( orange), $10^{6}$~(green), $10^{7}$~( red) solar masses. Right:  azimuthal shift of the cloud position from the initial  azimuth angle according to~Eq.~\ref{eq::ashift}. }\label{fig::dist_phase} 
\end{figure*}

For the general case of massive cloud motion the Chandrasekhar' dynamical friction formula can be written in the form:
\begin{equation}
\Oo \beta =  \frac{16 \pi^2 \ln \Lambda m_c(m_c+m*)}{v^3} \int\limits^{\bf v}_0 f({\bf r},{\bf u}){\bf 
u^2} d{\bf u}\,.
\end{equation}
where $f({\bf r},{\bf u})$ is the phase-space distribution function, $m*$  is the mass of a star, $m_c$ is the mass of a cloud. If the low mass particles are distributed in the disc with exponentially decreasing radial profile of the density, then the drag force is proportional to the coefficient~\citep{2004MNRAS.351.1215B}:
\begin{equation} 
\Oo \beta_d = \frac{3 \pi G \ln \Lambda \rho_0 m^2_c}{(\sqrt{2}\sigma)^3}\,,
\end{equation}
where $\rho_0$ is the local disc volume density and $\sigma$ is the velocity dispersion of stars.
 
For the case of Plummer sphere, representing the stellar spherical component,  the drag coefficient may be written in the following form~\citep{2002ApJ...572..371C}:
\begin{equation}
\Oo \beta_b = \frac{128\sqrt{2}}{7 \pi} \ln \Lambda \left( \frac{G}{M_b a} 
\right)^{1/2}m^2_c\,,
\end{equation}
where $M_b$ is the mass of spherical component and $a$ is its scale length.

Following ~\cite{1984MNRAS.209..729T}, we take the Coulomb logarithm $\Lambda$  as
\begin{equation}
\Oo \Lambda = \frac{b_{max}}{b_{min}}\,,  
\end{equation} 
where $b_{min} = {\rm max}(Gm_{cl}/V^2_0, 2R_{\rm cl})$, $V_0$ is a typical cloud velocity with respect to stars, $R_{\rm cl}$ is the cloud radius. Distance interval, $b_{max}$ is taken as the radius of gravitational influence of a given cloud. In the  calculations we assume $b_{max} = 2$~kpc.

\section{Calculations}
We studied the dynamics of clouds with masses in the range $10^4-10^7 M_{\odot}$ in the inner part of M~33, consisting of the disk and spherical components. We also took into account a finite size of clouds, assuming that they have a constant surface density, so that a cloud linear size is linked with its mass by the simple relation
\begin{equation}
 m_{cl} = 200 R^2_{cl}\,
\end{equation}
according to the  third Larson's law~\citep{1981MNRAS.194..809L}, which was obtained for molecular clouds of the Milky Way and  is applicable to nearby galaxies. The size of clouds was used for calculation of the logarithmic term in the dynamic force expression.  Initial orbital velocities of the clouds were taken as the circular velocities at a given radius corresponding to the gravitational potential model for a given circular velocity curve~(see Fig.~\ref{fig::rotation}). The motion of GMCs was described by Eq.~\ref{eq::dyn1} for clouds with the constant mass in the range $10^4-10^{7}$ solar masses ($M_\odot$)  during the time interval up to $5\times10^8$~yr. Initial radial coordinates were chosen as  $r_0 = 500\,, 1000\,, 1500\,, 2000$~pc.  

Although the HI and CO-maps of M~33 reveal a spiral-like structure,  the  Grand Design wave-like nature of spiral  arms in this galaxy remains very controversial: see ~\citet{1993AJ....105..499R,1990ApJ...349..497C,2007ApJ...661..830R}. Non-direct measurements of the corotation radius under the assumption of rigid-body rotation of spiral arms carried out by different authors led to a large spread of the estimates between 2.9 kpc and 5.9 kpc~\citep{2013MNRAS.428..625S}, which corresponds to the angular velocity of spiral pattern $\Omega_p$  between 18 and 33~\kmskpc.  Selecting the middle value of $\Omega_p$ = 25~\kmskpc, we show in~Fig.~\ref{fig::evol} the expected shift of  GMCs and spiral pattern in the differentially rotating disc as it is seen in the fixed reference coordinate system (the effect of dynamic friction is included).  The initial angular velocity of every cloud corresponds to the observed rotation curve of M~33 at a given initial radius $r_0$. The azimuth angle initially taken as $\varphi$ changes as :

\begin{equation}\label{eq::ashift}
\delta\varphi = \varphi - \Omega_0 (r_0)t
\end{equation} 
where $\varphi$ is a true polar coordinate of a cloud and $\Oo \Omega_0(r_0) = 
\frac{V_c(r_0)}{r_0}$  is the initial angular cloud speed. Fig.\ref{fig::evol} clearly illustrates that the lifetime of GMCs  should be very short (about $10^7$ yr) to allow them to remain linked with spiral arms.  Otherwise the observed connection between GMCs and the spiral like gaseous structure gives evidence of flocculent (``material")  nature of spiral arm in the inner several kpc-region of M~33.

\begin{figure}
\includegraphics[width=1\hsize]{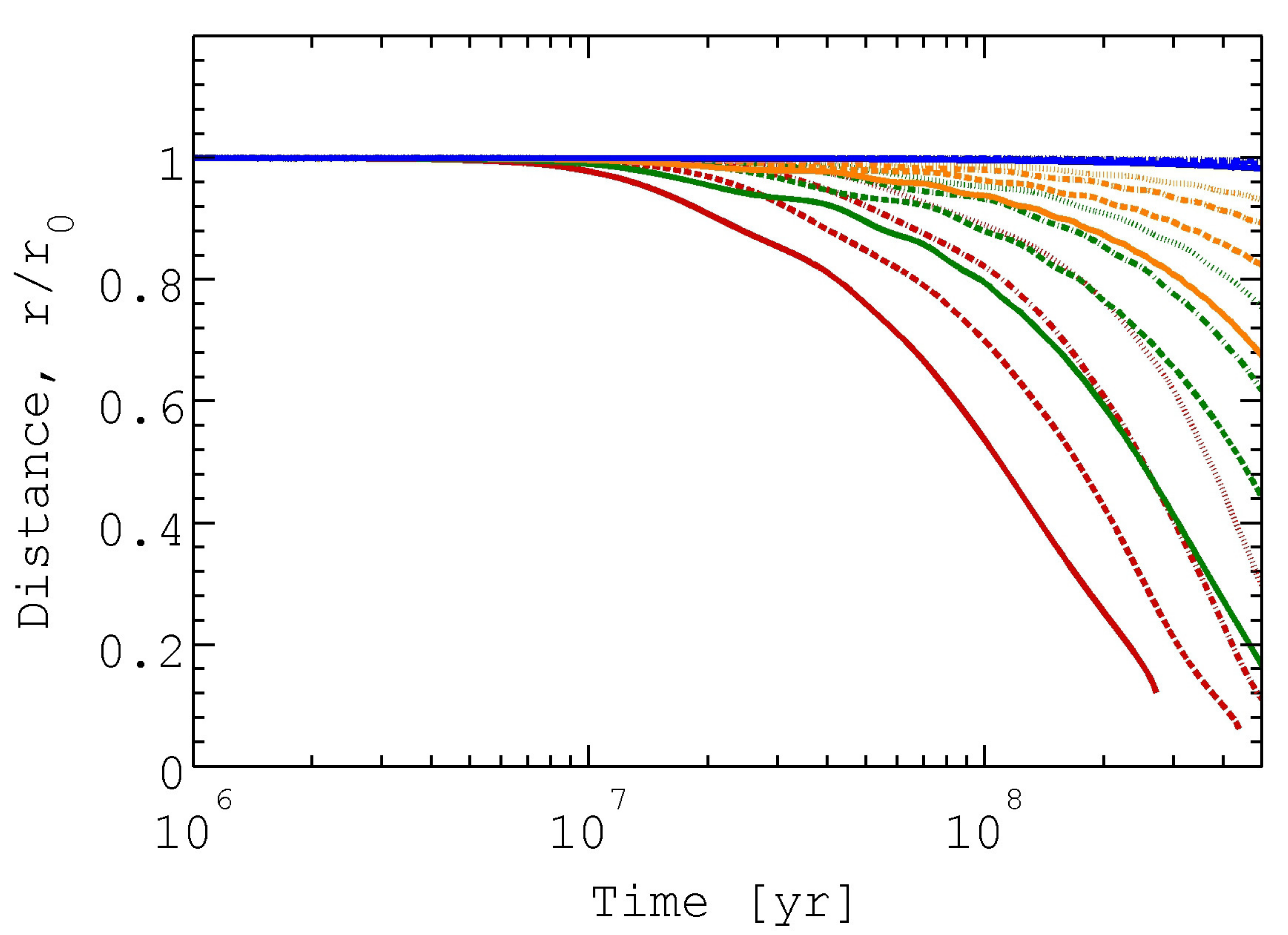}
\caption{Evolution of the galactocentric distances of GMCs normalized to the initial distances. Notations are the same as in {\bf Fig.~\ref{fig::dist_phase}}.\label{fig::reldyn}} 
\end{figure}

During the first $10$~Myr after their formation all clouds practically retain their initial circular orbits. After that their orbital radii decrease due to the loss of angular momentum by  dynamical friction. As one can expect,  more massive clouds fall down to the center much faster than the less massive ones. Note that the  clouds of equal masses starting   from the larger distances change their orbital radii faster than the clouds formed closer to the centre~(see Fig.~\ref{fig::reldyn}). After $\approx (1-3)\times 10^8$~yr the azimuthal shift with respect to the initial azimuth angle for massive clouds reaches $180-360^\circ$ (see Fig.~\ref{fig::dist_phase}). {\bf Fig.~\ref{fig::reldyn}} illustrates  the variation of radial distances of clouds normalized to the initial radius $r_0$. One can see that the relative change of radial coordinate depends both on the initial distance and a cloud mass. The increasing of drag force at lower radial distances  is the result of the increasing of the disc density and of the growing influence of spherical bulge component. 

\begin{figure}
\includegraphics[width=1\hsize]{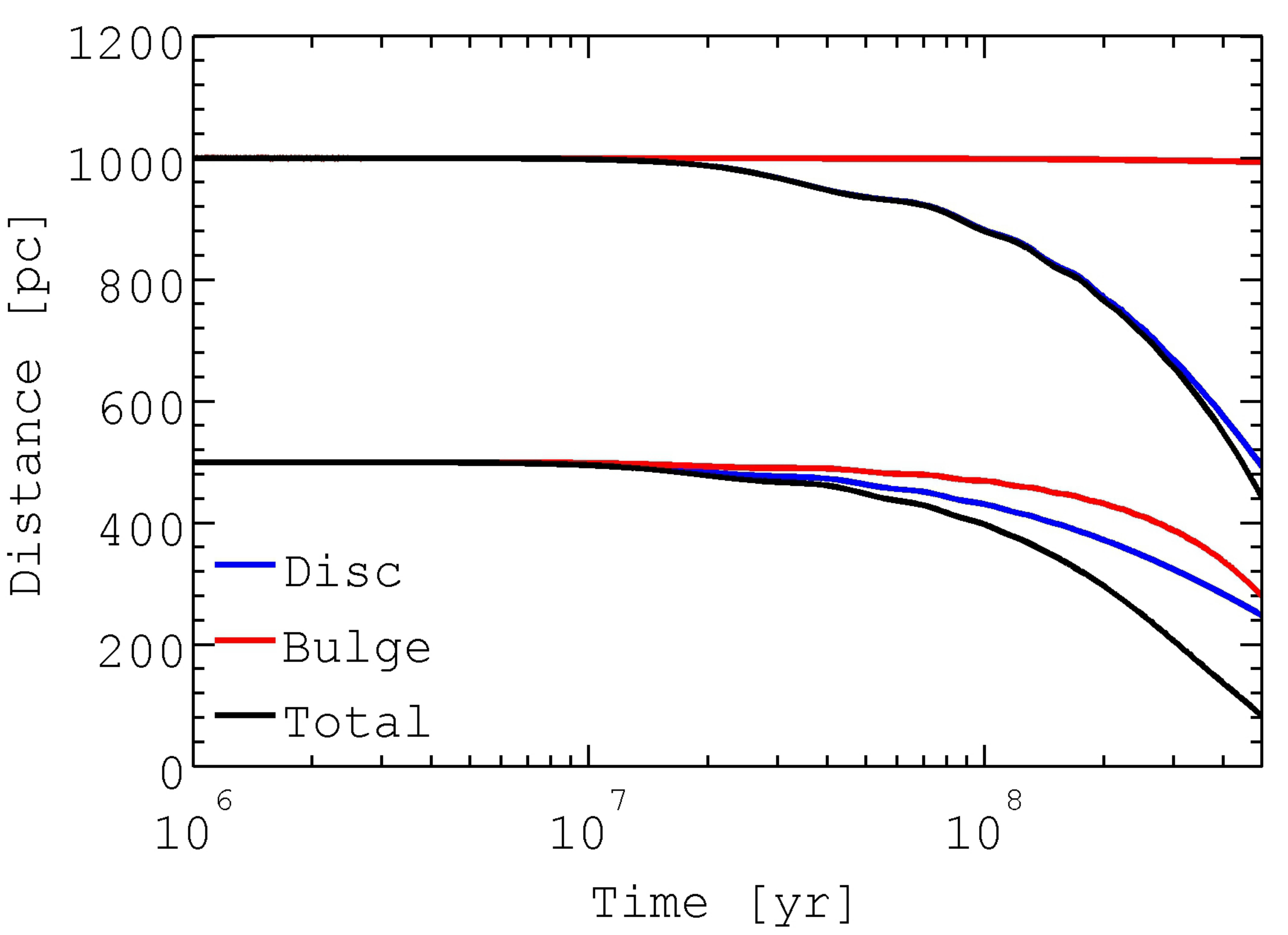}
\caption{Evolution of galactocentric distance of a cloud with the mass~$10^{7}$\Msun\, in the models where the drag is due to the bulge~(red), or the disc ~(blue) only, and the combined action of both components~(black). The latter case was illutrated in Fig.~\ref{fig::dist_phase}.} \label{fig::impact}
\end{figure}

To check the relative role of the components we calculate GMCs drift assuming $\beta_d = 0$ (the disc is absent) or $\beta_b = 0$ (the bulge is absent). As an illustration, in Fig.~\ref{fig::impact} we trace the dynamics of massive ( $10^7$~\Msun) cloud for the initial distances 1000 and 500~pc. For a cloud starting far away from the center the stellar bulge impact in the total radial shift is incredibly small. However in the central 500~pc the bulge plays a significant role (up to 30\%) due to its enhanced density, however the stellar disc remains the dominant factor of the clouds migration there.

A significant fraction of GMCs in M~33 is observed between radial distances $r = 2 - 4$~kpc.  A major amount of them have $M<3\cdot10^5M_\odot$~\citep{2007ApJ...661..830R}. Most of these GMCs are evidently situated near  their birthplaces inside of  the not-too-regular gaseous spiral-like filaments. We have not calculate the drag force for these outer clouds to clarify how their position may  changed during $\approx 10^8$ yr of their lifetimes, however the obtained results for $r\approx 2$~kpc demonstrate that the dynamic friction does not shift them significantly from their nearly circular orbits. Indeed, for the cloud masses $10^5-10^6 M_\odot$ at the initial distant $2$~kpc the change of radial distance is about $100-150$~pc~($<40''$) during $10^8$~yr after their formation (see Fig.~\ref{fig::dist_phase}), which is comparable with the typical width of HI filaments they are embedded. The mass function of GMCs in M~33 is very steep: $dN ~ M^{-2} dM$~\citep{2007ApJ...661..830R} with the mass truncation at $~10^6 M_\odot$, hence the radial shift due to dynamic friction may be noticeable only for a few most massive GMCs at $r  > 4$~kpc --- and only if they remain non-disrupted for that time interval. Note however that the  shift between GMCs and spiral arms would be significant even after much shorter time interval $~10^7$~yr, if the angular velocity of spiral pattern and of the disc in a given galactocentric interval were different, as it is illustrated in Fig~2. We consider this circumstance as the argument in favor of flocculent  nature of gaseous filaments, leaving the question of wave-like nature of  the most prominent stellar arms in M~33 open.

\section{Conclusions} 
As it was demonstrated above,  GMCs we considered in our model (with the exception of the most distant and less massive ones) change significantly their positions after~$10^8$ yr due to dynamical friction. The key role in this process is played by the  disk.

We may conclude that the dynamical friction may naturally explain the loss of connection between GMCs and spiral-like gaseous filaments in the inner 2 kpc in M~33. In turn it allows us to avoid the assumption that GMCs observed in this region were formed  beyond the gas filaments in contrast to the more distant counterparts.  If this approach is correct, it leads to three important conclusions.
\begin{itemize}
\item	Lifetimes of GMCs (at least in the inner region of M~33) should be close to or 
exceed $~10^8$ yr for the friction to be efficient ~\citep[see the discussion of the long lifetime of molecular clouds in][]{2014Ap&SS.353..595Z}.  

\item Spiral-like gaseous arms containing GMCs should also be long-lived 
features, otherwise the observed  GMCs would not be tightly linked to the arms at the distances  $r\geq$ 2 kpc.

\item Gaseous spiral arms in M~33 do not rotate rigidly, otherwise a significant fraction of GMCs would have observed in the inter-arm space (see Fig~\ref{fig::evol}). As long as the positions of most GMCs coincide with gaseous arms at radial distances 2-4 kpc, one may conclude that they move together with the gas, that is the gaseous arms are not the wave-like phenomena. Note that the flocculent nature of spiral arms in M~33 was argued earlier by different authors ~\citep[see e.g.][]{1990ApJ...349..497C,1993AJ....105..499R,2007ApJ...661..830R}. This conclusion is correct at least within the galactocentric radius $r \approx 4$~kpc: beyond that distance GMCs are nearly absent in spite of the existence of molecular gas~\citep{2007ApJ...661..830R}.

\end{itemize}

\section{Acknowledges}
We thank the anonymous referee for  fruitful comments. We also thank Evgenii Polyachenko for reading the manuscript and several helpful comments. The numerical simulations have been performed at the Research Computing Center~(Moscow State University) and Joint Supercomputer Center~(Russian Academy of Sciences). The numerical simulation, carried out at the Research Computing Center of the Lomonosov' Moscow state university,  was supported by the NSF  grant no. 14-22-00041. One of the authors (SAK) has been supported by a postdoctoral fellowship sponsored by the Italian MIUR. The work was also partially supported by the President RF grant (MK-4536.2015.2), and RFBR grants (15-02-06204, 15-32-21062).

\bibliography{m33bib}

\begin{thebibliography}{}

\bibitem[\protect\citeauthoryear{{Benson}, {Lacey}, {Frenk}, {Baugh} \&
  {Cole}}{{Benson} et~al.}{2004}]{2004MNRAS.351.1215B}
{Benson} A.~J.,  {Lacey} C.~G.,  {Frenk} C.~S.,  {Baugh} C.~M.,    {Cole} S.,
  2004, \mnras, 351, 1215

\bibitem[\protect\citeauthoryear{{Cepa} \& {Beckman}}{{Cepa} \&
  {Beckman}}{1990}]{1990ApJ...349..497C}
{Cepa} J.,  {Beckman} J.~E.,  1990, \apj, 349, 497

\bibitem[\protect\citeauthoryear{{Chatterjee}, {Hernquist} \&
  {Loeb}}{{Chatterjee} et~al.}{2002}]{2002ApJ...572..371C}
{Chatterjee} P.,  {Hernquist} L.,    {Loeb} A.,  2002, \apj, 572, 371

\bibitem[\protect\citeauthoryear{{Corbelli}}{{Corbelli}}{2003}]{2003MNRAS.342..199C}
{Corbelli} E.,  2003, \mnras, 342, 199

\bibitem[\protect\citeauthoryear{{Corbelli}, {Thilker}, {Zibetti}, {Giovanardi}
  \& {Salucci}}{{Corbelli} et~al.}{2014}]{2014A&A...572A..23C}
{Corbelli} E.,  {Thilker} D.,  {Zibetti} S.,  {Giovanardi} C.,    {Salucci} P.,
   2014, \aap, 572, A23

\bibitem[\protect\citeauthoryear{{Corbelli} \& {Walterbos}}{{Corbelli} \&
  {Walterbos}}{2007}]{2007ApJ...669..315C}
{Corbelli} E.,  {Walterbos} R.~A.~M.,  2007, \apj, 669, 315

\bibitem[\protect\citeauthoryear{{Dobbs} \& {Pringle}}{{Dobbs} \&
  {Pringle}}{2013}]{2013MNRAS.432..653D}
{Dobbs} C.~L.,  {Pringle} J.~E.,  2013, \mnras, 432, 653

\bibitem[\protect\citeauthoryear{{Dobbs}, {Pringle} \& {Naylor}}{{Dobbs}
  et~al.}{2014}]{2014MNRAS.437L..31D}
{Dobbs} C.~L.,  {Pringle} J.~E.,    {Naylor} T.,  2014, \mnras, 437, L31

\bibitem[\protect\citeauthoryear{{Jogee}, {Kenney} \& {Smith}}{{Jogee}
  et~al.}{1999}]{1999ApJ...526..665J}
{Jogee} S.,  {Kenney} J.~D.~P.,    {Smith} B.~J.,  1999, \apj, 526, 665

\bibitem[\protect\citeauthoryear{{Koda}}{{Koda}}{2013}]{2013ASPC..476...49K}
{Koda} J.,  2013, in {Kawabe} R.,  {Kuno} N.,   {Yamamoto} S.,  eds, New Trends
  in Radio Astronomy in the ALMA Era: The 30th Anniversary of Nobeyama Radio
  Observatory Vol.~476 of Astronomical Society of the Pacific Conference
  Series, {Evolution of Giant Molecular Clouds in Nearby Galaxies}.
p.~49

\bibitem[\protect\citeauthoryear{{Larson}}{{Larson}}{1981}]{1981MNRAS.194..809L}
{Larson} R.~B.,  1981, \mnras, 194, 809

\bibitem[\protect\citeauthoryear{{Miura}, {Kohno}, {Tosaki}, {Espada}, {Hwang},
  {Kuno}, {Okumura}, {Hirota}, {Muraoka}, {Onodera}, {Minamidani}, {Komugi},
  {Nakanishi}, {Sawada}, {Kaneko} \& {Kawabe}}{{Miura}
  et~al.}{2012}]{2012ApJ...761...37M}
{Miura} R.~E.,  {Kohno} K.,  {Tosaki} T.,  {Espada} D.,  {Hwang} N.,  {Kuno}
  N.,  {Okumura} S.~K.,  {Hirota} A.,  {Muraoka} K.,  {Onodera} S.,
  {Minamidani} T.,  {Komugi} S.,  {Nakanishi} K.,  {Sawada} T.,  {Kaneko} H.,
   {Kawabe} R.,  2012, \apj, 761, 37

\bibitem[\protect\citeauthoryear{{Murray}}{{Murray}}{2011}]{2011ApJ...729..133M}
{Murray} N.,  2011, \apj, 729, 133

\bibitem[\protect\citeauthoryear{{Regan} \& {Vogel}}{{Regan} \&
  {Vogel}}{1994}]{1994ApJ...434..536R}
{Regan} M.~W.,  {Vogel} S.~N.,  1994, \apj, 434, 536

\bibitem[\protect\citeauthoryear{{Regan} \& {Wilson}}{{Regan} \&
  {Wilson}}{1993}]{1993AJ....105..499R}
{Regan} M.~W.,  {Wilson} C.~D.,  1993, \aj, 105, 499

\bibitem[\protect\citeauthoryear{{Rosolowsky}, {Keto}, {Matsushita} \&
  {Willner}}{{Rosolowsky} et~al.}{2007}]{2007ApJ...661..830R}
{Rosolowsky} E.,  {Keto} E.,  {Matsushita} S.,    {Willner} S.~P.,  2007, \apj,
  661, 830

\bibitem[\protect\citeauthoryear{{Saburova} \& {Zasov}}{{Saburova} \&
  {Zasov}}{2012}]{2012AstL...38..139S}
{Saburova} A.~S.,  {Zasov} A.~V.,  2012, Astronomy Letters, 38, 139

\bibitem[\protect\citeauthoryear{{Saburova} \& {Zasov}}{{Saburova} \&
  {Zasov}}{2013}]{2013AN....334..785S}
{Saburova} A.~S.,  {Zasov} A.~V.,  2013, Astronomische Nachrichten, 334, 785

\bibitem[\protect\citeauthoryear{{Scarano} \& {L{\'e}pine}}{{Scarano} \&
  {L{\'e}pine}}{2013}]{2013MNRAS.428..625S}
{Scarano} S.,  {L{\'e}pine} J.~R.~D.,  2013, \mnras, 428, 625

\bibitem[\protect\citeauthoryear{{Scoville}}{{Scoville}}{2013}]{2013seg..book..491S}
{Scoville} N.~Z.,  2013, {Evolution of star formation and gas}

\bibitem[\protect\citeauthoryear{{Sil'chenko} \& {Lipunov}}{{Sil'chenko} \&
  {Lipunov}}{1987}]{1987Afz....26..443S}
{Sil'chenko} O.~K.,  {Lipunov} V.~M.,  1987, Astrofizika, 26, 443

\bibitem[\protect\citeauthoryear{{Tasker}, {Wadsley} \& {Pudritz}}{{Tasker}
  et~al.}{2015}]{2015ApJ...801...33T}
{Tasker} E.~J.,  {Wadsley} J.,    {Pudritz} R.,  2015, \apj, 801, 33

\bibitem[\protect\citeauthoryear{{Tosaki}, {Kuno}, {Onodera} Rie, {Sawada},
  {Muraoka}, {Nakanishi}, {Komugi}, {Nakanishi}, {Kaneko}, {Hirota}, {Kohno} \&
  {Kawabe}}{{Tosaki} et~al.}{2011}]{2011PASJ...63.1171T}
{Tosaki} T.,  {Kuno} N.,  {Onodera} Rie S.~M.,  {Sawada} T.,  {Muraoka} K.,
  {Nakanishi} K.,  {Komugi} S.,  {Nakanishi} H.,  {Kaneko} H.,  {Hirota} A.,
  {Kohno} K.,    {Kawabe} R.,  2011, \pasj, 63, 1171

\bibitem[\protect\citeauthoryear{{Tremaine} \& {Weinberg}}{{Tremaine} \&
  {Weinberg}}{1984}]{1984MNRAS.209..729T}
{Tremaine} S.,  {Weinberg} M.~D.,  1984, \mnras, 209, 729

\bibitem[\protect\citeauthoryear{{Williamson}, {Thacker}, {Wurster} \&
  {Gibson}}{{Williamson} et~al.}{2014}]{2014MNRAS.442.3674W}
{Williamson} D.~J.,  {Thacker} R.~J.,  {Wurster} J.,    {Gibson} B.~K.,  2014,
  \mnras, 442, 3674

\bibitem[\protect\citeauthoryear{{Yasutomi} \& {Tatematsu}}{{Yasutomi} \&
  {Tatematsu}}{1990}]{1990PASJ...42..517Y}
{Yasutomi} M.,  {Tatematsu} Y.,  1990, \pasj, 42, 517

\bibitem[\protect\citeauthoryear{{Zasov} \& {Kasparova}}{{Zasov} \&
  {Kasparova}}{2014}]{2014Ap&SS.353..595Z}
{Zasov} A.,  {Kasparova} A.,  2014, \apss, 353, 595

\end{thebibliography}
\end{document}